\documentclass[preprint]{elsarticle}

\usepackage{graphicx}

\newcommand{\cat}{+\!\!\!+}

\newtheorem{definition}{Definition}
\begin{document}

\title{{\bf \em Tensors and n-d Arrays:\\ A Mathematics of Arrays (MoA), 
\mbox{\boldmath $\psi$-}Calculus
and the Composition of Tensor and Array Operations
\footnote{Presented at the NSF Workshop: {\em On Future Directions in 
Tensor-Based Computation and Modeling}, February 20-21, 2009,  
NSF Arlington, VA 22230}
}\\
}

\author{Lenore M. Mullin} 
\address{National Science Foundation, Arlington, VA}
\author{James E. Raynolds}
\address{College of Nanoscale Science and Engineering, University
at Albany, State University of New York, Albany, NY}


\begin{abstract}
The Kronecker product is a key algorithm and is ubiquitous across the 
physical, biological, and computation social sciences. 
Thus considerations of optimal implementation are important. 
The need to have high performance and computational reproducibility is 
paramount.  Moreover, due to the need to compose multiple Kronecker products, 
issues related to data structures, layout and indexing algebra require
a new look at an old problem. This paper discusses the outer product/tensor 
product and a special case of the tensor product: the Kronecker product, 
along with optimal implementation when composed, and 
mapped to complex processor/memory hierarchies. We discuss how the use 
of {\em A Mathematics of Arrays} (MoA), and the $\psi$ {\em - Calculus},  
(a calculus of indexing with shapes), provides optimal, verifiable, 
reproducible, scalable, and portable implementations of both hardware and 
software~\cite{mul88,mul91,Mul03,mullin-raynolds-book}.
\end{abstract}
\setcounter{footnote}{1}

\begin{keyword}
Kronecker product \sep tensor product \sep tensor decomposition 
\sep processor/memory hierarchy \sep program optimization \sep matrix and 
array languages \sep multi-linear algebra \sep Mathematics of Arrays \sep
Conformal Computing\footnote{The name Conformal Computing \copyright $\;$ is
protected.  Copyright 2003, The Research Foundation of State University of 
New York, University at Albany.}

\MSC 15A63 \sep 15A69 \sep 20K25 \sep 46A32 \sep 46M05 \sep 47L05 \sep 47L20
\sep  46A32 \sep 46B28 \sep 47A80 
\end{keyword}

\maketitle

\newpage
\section{Introduction}
The purpose of this report is to discuss the outer product/tensor product and a special case of the
tensor product: the Kronecker product, as well as algorithms, their origin, 
and optimal implementation when composed, and mapped to complex processor/memory hierarchies~\cite{vanloan00,graham,davio1981,horn91,steeb97,henderson83,
henderson81,snay78,regalia89}. We discuss how the use of {\em A Mathematics of Arrays} (MoA), and the $\psi$ {\em - Calculus},  (a calculus of indexing with 
shapes), provides optimal, verifiable, reproducible, scalable, and portable implementations of both hardware and software~\cite{mul88,mul91,Mul03,mullin-raynolds-book}.
This is due to the fact that we are using normal forms composed of multi-linear operations on Cartesian coordinates which are transformed into simple abstract machines:  {\bf starts, stops, strides, count}, up and
down the processor/memory hierarchy.  Before turning the the discussion at 
hand we invite the reader to consult the appendix for motivational background
illustrating how tensor/Kronecker products and diadics arise naturally in 
applied problems in physics and engineering.

A key notion of the present work is how the MoA outer product can be formulated as the Kronecker product, a special case of the Tensor product.  We will show
that the use of the MoA outer product is superior to the traditional
approach when one is concerned with efficient implementations of multiple
Kronecker products.
The MoA outer product is a general operation on two arrays of any
shape or dimension and applies any scalar operation, not just the 
product ($*$) on these two arrays (i.e. $+$, $-$, $/$, etc. are 
valid operations). 
For now, we focus on the relationship to the Kronecker product between two matrices of arbitrary size resulting in a block matrix.
Let's begin with an example where:
\[ A = 
\left[
\begin{array}{cc}
1  & 2    \\
 3 & 4    \\
\end{array}
\right],
\mbox{ and }
B=\left[
\begin{array}{cccc}
  5& 6  &7 &8  \\
  9& 10 & 11 &12\\
 13 &14&15 &16
\end{array}
\right],
\]
the operation:
\[ A \bigotimes B, \]
is defined by the operation in Fig.~\ref{kprodab}.
\begin{figure}[h]
\[
\left[
\begin{array}{cc}
1  & 2    \\
 3 & 4    \\
\end{array}
\right]
\bigotimes
\left[
\begin{array}{cccc}
  5& 6  &7 &8  \\
  9& 10 & 11 &12\\
 13 &14&15 &16
\end{array}
\right]
=
\]
\[
\left[
\begin{array}{llllllll}
 1\times5 & 1\times 6 &1\times7 &1\times 8&2\times 5&2\times 6&2\times7&2\times 8 \\
 1\times9 & 1\times10  &1\times11 &1\times 12&2\times 9&2\times 10&2\times11&2\times 12 \\
1\times13 & 1\times 14  &1\times15 &1\times 16&2\times 13&2\times 14&2\times15&2\times 16 \\
3\times5 & 3\times 6  & 3\times7 &3\times 8&4\times 5&4\times 6&4\times7&4\times 8  \\
 3\times9 & 3\times10  & 3\times11 &3\times 12&4\times 9&4\times 10&4\times11&4\times 12 \\
3\times13 & 3\times 14  & 3\times15 &3\times 16&4\times 13&4\times 14&4\times15&4\times 16 \\
\end{array}
\right]
\]
\caption{Kronecker product of A and B: A matrix.}
\label{kprodab}
\end{figure}

Note implicitly in the operation above, that  the 4 multiplications applied to B have a substructure within
the resultant array.  That is, EACH component of A is multiplied with ALL of B creating 4,  $3 \times 4$ arrays. 
The result is stored in a matrix, C, by relating the indices of A, i.e. i,j, with the indices of B, i.e. k,l.  and encoding them into row, column coordinates. Classically, i,k is correlated to a row, and j,l is correlated to a column. More on this later. 

A goal of this paper is to describe how {\bf \em shapes} are integral to 
array/tensor operations.  By definition, the {\bf \em shape} of an array 
is a vector whose elements equal the length of each corresponding dimension 
of the array.  Using shapes, we will relate operations in A Mathematics of Arrays (MoA) to tensor algebra and we will show how these shapes and the 
$\psi$-Calculus (also sometimes written: Psi-calculus) can be used to compose multiple Kronecker products and map such operations to complex processor/memory hierarchies. .
\section{Shapes and the $\psi$ operator}

Let's begin by introducing shapes. The shape of A is 2 by 2, i.e. 
$\rho A =\; <2\;2>$, the shape of B is 3 by 4, i.e. $\rho B = \;<3\;4>$ and the shape of
$A \bigotimes B$ is 6 by 8, i.e. $\rho ( A \bigotimes B )= \;<6\;8>$.  In this
discussion we have introduced the {\em shape} operator, $\rho$, which acts on an array and
returns its shape vector.

Now, let's look at the MoA outer product of A and B, denoted by 
$A\; {op}_\times B$. The shape of $A\; {op}_\times B$ is the concatenation of the shapes of A and B, i.e.
a 4-dimensional array with shape 2 x 2 x 3 x 4. 
That is, $\rho \;( A\; {op}_\times B ) = \;\; <2\;2\;3\;4>$.  The resulting 
array is indexed by a vector $<\!i\,j\,k\,\ell\!>$ that is ordered 
in row-major order (i.e. in the order of a nested $\{\,i\,j\,k\,\ell\,\}$ 
loop with $\ell$ the fastest and $i$ the slowest increasing {\em partial 
index}).

\begin{figure}[h]
\[
\left[
\begin{array}{cc}
1  & 2    \\
 3 & 4    \\
\end{array}
\right]
{op}_\times 
\left[
\begin{array}{cccc}
  5& 6  &7 &8  \\
  9& 10 & 11 &12\\
 13 &14&15 &16
\end{array}
\right]
=
\]
\[
\left [
\begin{array}{cc}
\left[
\left[
\begin{array}{cccc}
 1\times5 & 1\times 6 &1\times7 &1\times 8 \\
 1\times9 & 1\times10  &1\times11 &1\times 12\\
1\times13 & 1\times 14  &1\times15 &1\times 16
\end{array}
\right]
\left [
\begin{array}{cccc}
2\times 5&2\times 6&2\times7&2\times 8 \\
 2\times 9&2\times 10&2\times11&2\times 12 \\
2\times 13&2\times 14&2\times15&2\times 16 \\
\end{array}
\right]
\right ] 
\\

\left[
\left[
\begin{array}{cccc}
 3\times5 & 3\times 6  & 3\times7 &3\times 8  \\
 3\times9 & 3\times10  & 3\times11 &3\times 12 \\
3\times13 & 3\times 14  & 3\times15 &3\times 16
\end{array}
\right]
\left[
\begin{array}{cccc}
4 \times 5&4\times 6&4\times7&4\times 8  \\
 4\times 9&4\times 10&4\times11&4\times 12 \\
4\times 13&4\times 14&4\times15&4\times 16 \\
\end{array}
\right]
\right]
\end{array}
\right]
\]
\caption{\label{moaproduct}MoA Outer product of A and B: A 4-d array}
\end{figure}
The MoA array operation: $A\; {op}_\times B$ is defined by the result
in Fig.~\ref{moaproduct}.

Notice that the layouts in Figs.~\ref{kprodab}  and~\ref{moaproduct}  are very similar. What is different is the bracketing.
The result of the MoA outer product is NOT a matrix but is rather a 
multi-dimensional array. In contrast, the result of the Kronecker product IS
a matrix (i.e. a two-dimensional array). The extra brackets reflect the fact that the result of the outer product is a 4-dimensional array whose shape is obtained by concatenating the
 shapes of the arguments, i.e. $<2\;2>$ concatenated to $<3\;4>$ equals $<2\;2\;3\;4>$.
So do these arrays have the same layout in memory? The answer is no. What is interesting, however, is that
when the Kronecker product is executed it is filled in, in a row major ordering relative to
the right argument. 
The layout, either row or column major, would reflect the access patterns needed to optimize these operations across the processor/memory hierarchy. 
Let's assume row major. Thus flattening (i.e. creating a vector consisting of
the elements of the array in row-major order), the difference in layout is as follows:
\begin{figure}[h]

\begin{equation}
\left <
\begin{array}{llllllll}
 1\times5 & 1\times 6 &1\times7 &1\times 8&2\times 5&2\times 6&2\times7&2\times 8 \;\;.\;.\;.\\
\end{array}
\right >
\label{flatkron}
\end{equation}
\nonumber
\caption{Kronecker product flattened using row-major layout}
\end{figure}

\begin{figure}[h]
\begin{equation}
\left <
\begin{array}{llllllll}
 1\times5 & 1\times 6 &1\times7 &1\times 8 & 1\times9 & 1\times10  &1\times11 &1\times 12
 \;\;.\;.\;.\\
\end{array}
\right >
\label{flatmoa}
\end{equation}
\caption{MoA Outer product flattened using row-major layout}
\end{figure}

Before we continue, let's discuss how languages implement these operations. 
Typically, assuming A, B, and C are defined as n by n arrays, the operation:
\[ A \bigotimes {B \bigotimes C} \]
would materialize all of
$B \bigotimes C $ as a temporary array, let's call it TEMP. Then it would perform
$A \bigotimes  TEMP$. If n is large, this could use an enormous amount of space.

Now, let's look at how MoA and $\psi$-Calculus would perform the outer product. Then, we'll discuss how
we can restructure the MoA outer product to get the Kronecker product and in so doing we'll be able
to {\bf \em compose multiple Kronecker products efficiently and deterministically over complex processor/memory hierarchies.}

\subsection{Shapes and the Outer product}

Before beginning, we refer the reader to the numerous publications on MoA and the $\psi$-Calculus, the most foundational is given in Ref.~\citep{mul88}.  We thus take liberty to use operations in the algebra and calculus by example. Only when necessary will a definition be given.
\begin{definition}

Assume A, B, C, are $n\times n$ arrays, that is, each array has shape:  
\[\rho A = \rho B = \rho C = \;<\;n\;n\;>.\]
Assume the existence of the $\psi$ operator and that it is well defined 
for n-dimensional arrays.  The $\psi$ operator takes as left argument
an index vector and an array as the right argument and returns the 
corresponding component of the array.  For a full index (i.e. as many
components are there are dimensions) a scalar is returned and for 
a {\em partial index}, a sub-array is selected.
Then, 
\[  D = A\; {op}_\times ( B  \; {op}_\times C )                      \]
is defined when the shape of D is equal to the shape of $A\; {op}_\times ( B  \; {op}_\times C ) $.
And the shape of $A\; {op}_\times ( B  \; {op}_\times C ) $ is equal to the shape of
$A$ concatenated to the shape of $( B  \; {op}_\times C ) $ which is equivalent to
the shape of $A$ concatenated to the shape of $B$ concatenated to the shape of $C$. i.e.
\[ \rho D = \rho  (A\; {op}_\times ( B  \; {op}_\times C ) )  = \rho A \cat \rho ( B  \; {op}_\times C ) =
\rho A \cat \rho B \cat \rho C  = \;
<n\;n\;n\;n\;n\;n>\]
\noindent
Then,  $\forall\;\; i_0,j_0,k_0,l_0,m_0,n_0 \;s.t.$\\
$0 \leq i_0<n 
;\;0 \leq j_0<n
;\;0\leq k_0<n
;\;0\leq l_0<n
;\;0\leq m_0<n
;\;0\leq n_0< n$ 
\end{definition}
\begin{eqnarray}
<i_0\;j_0\;k_0\;l_0\;m_0\;n_0> \psi \;D &=& (<i_0\;j_0> \psi \; A) \times (<k_0\;l_0\;m_0\;n_0> \psi  \;( B  \; {op}_\times C ))\nonumber\\
&=&(<i_0\;j_0> \psi  \;A) \times ( <k_0\;\l_0> \psi  \;B) \times (<m_0\;n_0> \psi \; C ) \nonumber
\end{eqnarray}
It is easy to see that we can compose as little or as much as we like given 
the bounds of $i_0,j_0,k_0,l_0,m_0$ and $n_0$.
We'll return to how to build the above composition. We'll also discuss how to include processor memory hierarchies but first we'll discuss how to make the layout of the Kronecker product equivalent to the layout of the MoA outer product.

\subsection{Permuting the indices of the MoA outer product}
In order to discuss permuting the outer product we must first discuss how to permute an array. One way is through a transpose. We are familiar with transposing an array, i.e $A^T$. We know that $A[j;i]$ denotes $A^T[i,j]$. Let's now discuss how to  transpose a matrix in MoA and then how to transpose
an array in general. 
\begin{definition}
Given the shape of A is m by n,  i.e. $\rho A = <m\;n>$. then
$A^T$ is defined when the shape of $A^T $ is n by m. That is,
\[ \rho A^T = <\;n\;m>.\]
Then, for all $0 \leq i < n $ and $ 0 \leq j<m$
\[ <i\;j> \psi A^T = <j\;i> \psi A \]
\end{definition}
Let's now generalize this to any arbitrary array.
\begin{definition}
Given the shape of A is 
$<m\;n\;o\;p\;q\;r>$.
Then $A^T$ is defined when the shape of
$A^T$ is 
$<r\;q\;p\;o\;n\;m>$
Then for all $0 \leq i_0 < r$; $0 \leq j_0 < q$; $0 \leq k_0< p$; $0 \leq l_0 < o$; $0 \leq m_0 < n$; $0 \leq n_0< m$;
\[ <\!i_0\, j_0\, k_0\, l_0\, m_0\, n_0\! > \psi A^T = <\!n_0\, m_0\, l_0\, k_0\, j_0\, i_0\! > \psi A \]
\end{definition}
A question should immediately come to mind. Can the indices permute in other ways other than reversing them? The answer is yes, and in fact any permutation
consistent with the shape of the array is achieved by simply permuting the
elements of the index vector.  Note that the definitions for {\bf \em general 
transpose} and {\bf \em grade up} presented herein are the same definitions proposed to the F90 ANSI Standard Committee in 1993 and subsequently accepted for inclusion in F95.

\begin{definition}
The operator {\bf \em grade up} is defined for an $n$-element vector containing positive 
integers in the range from $0$ to $n-1$ in any order (multiple entries
of the same integer are allowed).  The result is a vector denoting the
positions of the lowest to the highest such that when the original vector is indexed by the result of grade up,  the original vector  is sorted from lowest to highest. \\
{\bf Example:} Given $\vec a\; =\; <2\; 0\; 1\; 3>, \;\; \;gradeup[\;\vec a\;] \;= \;gradeup[<2\;0\;1\;3>] \;= \;<1\;2\;0\;3>$. Thus, $\vec a[gradup[\vec a]] = \vec a[<1\;2\;0\;3>]=
<2\;0\;1\;3>[<1\;2\;0\;3>]\; = \;<0\;1\;2\;3>.$ 
\end{definition}
\noindent
To clarify this example we state the operations in words.  The $0$'th element
of the index vector is $1$, implying that the element in position $1$ of the
vector $\vec a$, i.e. $0$, should be placed in the $0$'th position of the 
result.  The $1$'st element of the index vector, $2$, implies that the $2$'nd
element of $\vec a$, i.e. $1$ should be placed in the $1$'st position of the
result and so on.
We are now ready to define a general transpose for n-dimensional arrays.
\begin{definition}
Given an array A with shape $\vec s$ such that the total number of components in $\vec s$ denotes
the dimensionality  d,  of A. $A^{T_{\vec t}}$ is defined whenever the shape of $A^{T_{\vec t}}$ is
$\vec s [\vec t]$, i.e. $\rho A^{T_{\vec t}}= \vec s [\vec t]$. 
Then, for all $0 \leq^* \vec i <^* \vec s [\vec t]$
(the symbols $\leq^*$ and $<^*$ imply element by element comparisons):
\[ \vec i \psi A^{T_{\vec t}} = \vec i[gradeup[\vec t]] \psi A\]
\end{definition}
{\bf Example:} 
Given  \[ A=
\left[
\left[
\begin{array}{ccc}
  0&1   &2   \\
  3& 4  &5   \\
 6 & 7  &8 \\
 9 & 10& 11  
\end{array}
\right]
\left[
\begin{array}{ccc}
  20&21   &22   \\
  23& 24  &25   \\
 26 & 27  &28 \\
 29 & 30& 31  
\end{array}
\right]
\right]
\]
We first look at
$A^{T_<2\;1\;0> }$ and note that this is equivalent to  $ A^T$. The shape of $A$ is $<2\;4\;3>$ so the
shape of $A^{T_<2\;1\;0> }$ is $<2\;4\;3>[<2\;1\;0>]= <3\;4\;2>$. 
Then for all \break $0 \leq^* \;\; <\!i\;j\;k\!>  \;\; <^* \;\; <\!3\;4\;2\!>$
(this is a shorthand notation for 
$0 \leq i < 3$; $0 \leq j < 4$; $0 \leq k < 2$ 
) we have:
\begin{eqnarray}
 <i\;j\;k> \psi A^{T_<2\;1\;0> } &=& (<i\;j\;k>[gradeup[<2\;1\;0>]] )\psi A \nonumber \\
 &=&( <i\;j\;k>[<2\;1\;0>]) \psi A \nonumber \\
 &=& <k\;j\;i> \psi A \\
 &=&
 \left[
\left[
\begin{array}{cc}
 0 & 20   \\
 3 & 23   \\
 6 & 26 \\
 9 & 29  
\end{array}
\right]
\left[
\begin{array}{cc}
 1   &21   \\
 4  &24   \\
 7  &27 \\
10& 30  
\end{array}
\right]
\left[
\begin{array}{cc}
  2   &22   \\
 5  &25   \\
 8  &28 \\
11& 31  
\end{array}
\right]
\right]
 \end{eqnarray}
\noindent 
Now let's look at another permutation of $A$ noting there are $6$ possible permutations, i.e.
$<0\;1\;2>, <0\;2\;1>, <1\;2\;0>, <1\;0\;2>,<2 \;0\;1>$, and $<2\;1\;0>$. 
\noindent
This time let's look at
$A^{T_{<2\;0\;1>}}$. Now the shape of $A^{T_{<2\;0\;1>}}$ is $<2\;4\;3>[<2\;0\;1>] = <3\;2\;4>$.
Then for all $0 \leq ^* \;\; <\!i\;j\;k\!> \;\; <^* \;\; <\!3\;2\;4\!>$
\begin{eqnarray}
 <i\;j\;k> \psi A^{T_{<2\;0\;1>}} &=& (<i\;j\;k>[gradeup[<2\;0\;1>] ]) \psi A \nonumber \\
&=& (<i\;j\;k>[<1\;2\;0>])  \psi A \nonumber \\
&=& <j\;k\;i> \psi A\\
&=&
\footnotesize
\left[
\left[
\begin{array}{cccc}
  0   &3 & 6&9 \\
  20& 23&26&29   \\
 \end{array}
\right]
\left[
\begin{array}{cccc}
 1   &4&7&9   \\
21  &24 &27&30  \\
 \end{array}
\right]
\left[
\begin{array}{cccc}
  2   &5&8&11   \\
22 &25 &28&31  \\
\end{array}
\right]
\right] \nonumber
\end{eqnarray}
\normalsize
\section{Changing Layouts using Permutations}

\begin{figure}[h]
\footnotesize
\[
\left [
\begin{array}{ccc}

\left[
\begin{array}{cccc}
 1\times5 & 1\times 6 &1\times7 &1\times 8 \\
 2\times 5&2\times 6&2\times7&2\times 8 \\
\end{array}
\right]

\left [
\begin{array}{cccc}
1\times9 & 1\times10  &1\times11 &1\times 12\\
 2\times 9&2\times 10&2\times11&2\times 12 \\
\end{array}
\right]

\left [
\begin{array}{cccc}
1\times13 & 1\times 14  &1\times15 &1\times 16\\
2\times 13&2\times 14&2\times15&2\times 16 \\
\end{array}
\right]

\\
\left[

\begin{array}{cccc}
3\times5 & 3\times 6  & 3\times7 &3\times 8  \\
4 \times 5&4\times 6&4\times7&4\times 8  \\
\end{array}
\right]

\left[
\begin{array}{cccc}
3\times9 & 3\times10  & 3\times11 &3\times 12 \\
 4\times 9&4\times 10&4\times11&4\times 12 \\
\end{array}
\right]

\left[
\begin{array}{cccc}
3\times13 & 3\times 14  & 3\times15 &3\times 16\\
4\times 13&4\times 14&4\times15&4\times 16 \\
\end{array}
\right]
\end{array}
\right ]
\]
\caption{\label{moaproductrans} Transpose of MoA Outer product of A and B: A 4-d array}
\end{figure}
Now that we know how to permute an array over any of it's dimensions we can reorient the MoA outer product to have the same layout as the Kronecker product or if we desire, we can reorient the Kronecker product to have the same layout as the MoA outer product. The pros and cons of each layout will be discussed in a later section.

Recall the layouts of the Kronecker product in Fig.~\ref{flatkron} and the MoA outer product in Fig.~\ref{flatmoa}. Let's first permute the MoA outer product such that it has the same layout as the Kronecker product, and study the 4-d array defined by the MoA outer product in Fig.~\ref{moaproduct}.
Now observe the array in Fig.~\ref{moaproductrans}. Flattening this 4-d array gives us the layout we want.
Notice which dimensions changed between the initial outer product in Fig.~\ref{moaproduct} and the transposed outer product in Fig.~\ref{moaproductrans}. The shape went from $2 \times 2 \times 3 \times 4$ to $2 \times 3 \times  2 \times 4$.  Reviewing equations \ref{flatkron} and \ref{flatmoa} we want
$1$ times $5,6,7,$ and $8$ to be next to $2$ times $5,6,7,$ and $8$, etc. in the layout. Thus, we want to leave the
0th dimension alone, the 3rd dimension alone and we wanted to permute the 1st dimension with the 2nd.  Consequently, we want  $ ( A\; {op}_\times B )^{T_{<0\;2\;1\;3>}}$, i.e. the $ <0\;2\;1\;3>$ transpose of the outer product of A and B.
Notice that this is the SAME permutation used in correlating the indices of A and B with the indices of the Kronecker product, i.e. resulting matrix, i.e. 
$i,j,k,l \rightarrow i,k,j,l$.

Recall that this is the same permutation we discussed for the transpose of the MoA outer product. We now can discuss how to optimize these computations. Using MoA and $\psi$ Calculus, one
can not only {\em compose} multiple indices in an array expression but, the algebraic
reformulation of an expression can  include processor/memory hierarchies. This
is done by increasing
the dimensions of the arguments. Through various restructurings, an expression
can easily describe how to scale and 
port across complex processor/memory architectures.

Unless familiar with the topic, see the Appendix which
gives a historical perspective of the Kronecker product and illustrates
how pervasive the inner and outer products are throughout 
science. That said, an efficient, correct, scalable, portable 
implementation becomes paramount, e.g.
accurate simulations and reproducible computational experiments rely on this.
 
History shows us how the resultant matrix of the Kronecker product is evaluated
and indexed. The permutations on the input matrices in conjunction with an equivalent 
permutation on the corresponding shapes followed by a pairwise multiplication determines
not only the resultant shape but how to store the results in its associated 
index  of the resultant array.  This cumbersome computation and encoding into 
new 2-d indices gets
more and more complicated as the number of successive Kronecker products increases.
Moreover, issues of parallelization complicate the problem since various components
in the left argument are used over the columns of the result, assuming the
partitioning was done by rows. Other partitions are possible: blocks, columns, etc..
When the input matrices are large the problem is further complicated. 
This is not the case in MoA and $\psi$ Calculus.

\section{Multiple Kronecker products}
Multiple Kronecker products are common in conjunction with inner
products and permutations such as transpose. How can these be optimized to use
basic abstract machine instructions at all levels up and down the 
processor/memory hierarchy: start, stop, stride, count?

Presently, multiple Kronecker products require the materialization of each pair of products. Notice what happens. After each pair of products, the result must
be stored using the permutations of the indices of the argument arrays and encoded into row/column coordinates in a new matrix with size equal to the product 
of the pairs of permuted shapes. For example, if the input arrays were 2 x 2 and 3 x 3. The resultant shape would be a (2 x 3) by (2 x 3), i.e. 6 x 6.
Now, if we then did a Kronecker product with a 2 x 2, the results would be
a 12 x 12. With each subsequent Kronecker product we'd need to  store
the product in the rows and columns associated with the permuted indices. 
Ideally, we want to {\em compose} multiple products in terms of their indexing.
MoA and $\psi$-calculus are ideally suited for this approach and easily facilitate not only the composition of multiple Kronecker/outer products but their mapping to complex processor memory hierarchies. 

\begin{figure*}[h]
\begin{center}
\includegraphics[width=3cm]{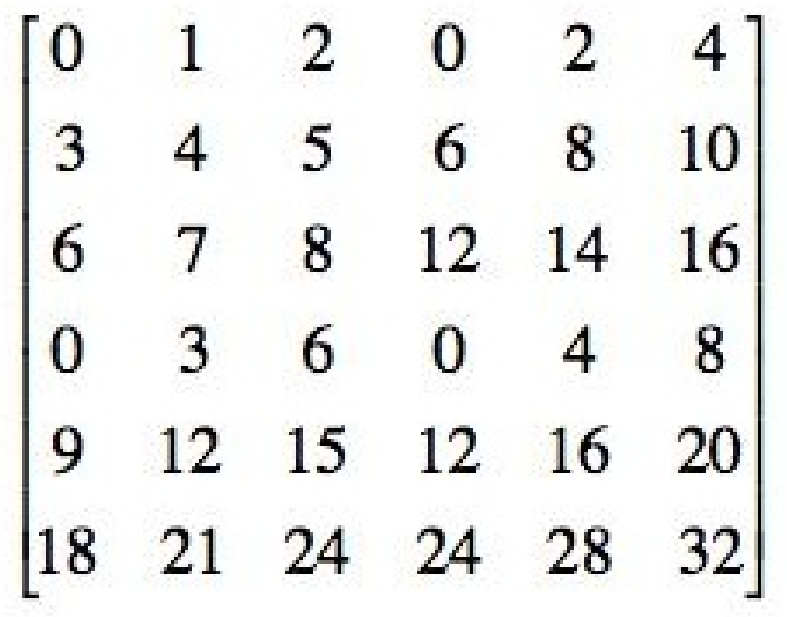}
\caption{\label{mat4}}
\end{center}
\end{figure*}

\begin{figure*}[h]
\begin{center}
\includegraphics[height=.5in,width=.75in]{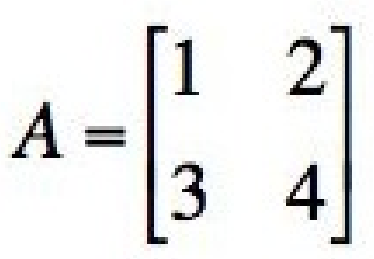}
\caption{\label{mat2}}
\end{center}
\end{figure*}

\begin{figure*}[h]
\begin{center}
\includegraphics[height=.5in,width=.75in]{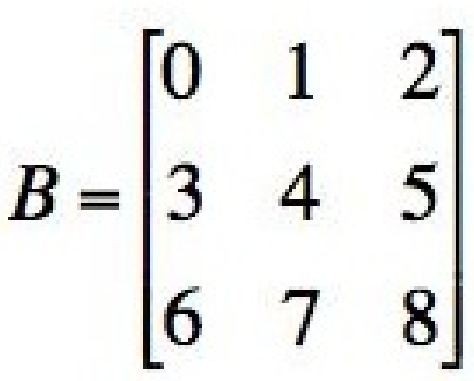}
\caption{\label{mat3}}
\end{center}
\end{figure*}

\begin{figure*}[h]
\begin{center}
\includegraphics[height=3.5in,width=3.75in]{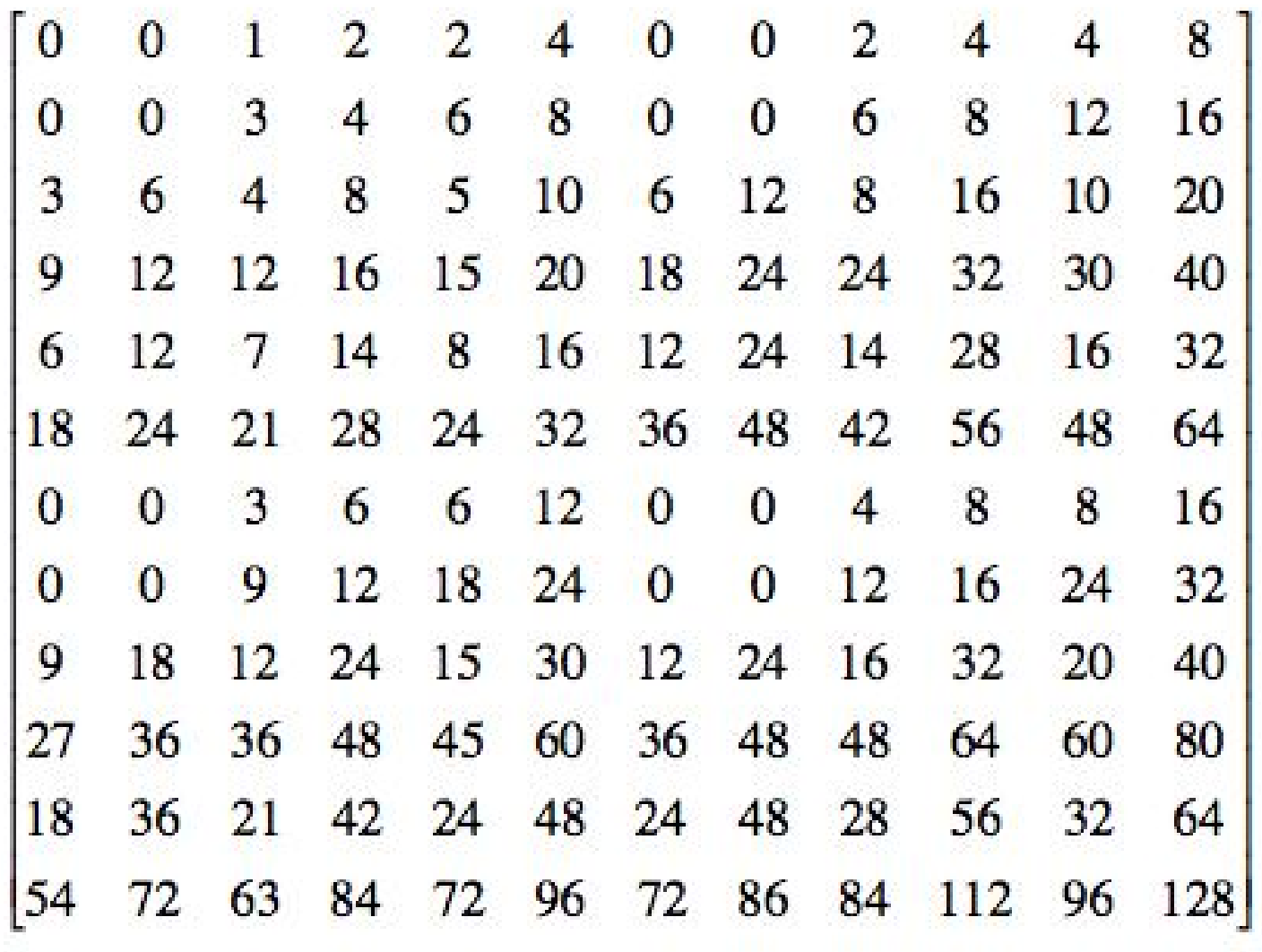}
\caption{\label{mat5}}
\end{center}
\end{figure*}

To illustrate,
let A be a 2 x 2 and B a 3 x 3 array. 
We are not concerned with the specific values of the matrix elements
since we need only to consider manipulations of the
indices. We assume the arithmetic is correctly defined. We'll perform 
$E = (A \bigotimes B) \bigotimes A $.
The result within the parentheses would have shape 6 x 6. This was due to the
two input  array shapes, i.e. 2 x 2 and 3 x 3. Using, $i$,$j$ in A and $k$,$l$ in B
bounded by their associated shapes, we combine $i$,$k$ with the associated shape
from that array, i.e. 2,2 and 3,3 are analogously permuted, then multiplied. 
Thus 2,3 and 2,3 become the new row, column associations. 
These are
then multiplied together to become the new number of rows and columns, i.e. 
new shape. The shapes above are used to {\em encode} the location
of each Kronecker product operation.
In other words, the composite index $i,k$ indexes the rows of 
$(A \bigotimes B)$ while the composite index $j,k$ indexes the 
columns of $(A \bigotimes B)$.
The resultant array, let's call it C, 
is shown in Figure~\ref{mat4}.
The input arrays are A and B (see Figures ~\ref{mat2} and ~\ref{mat3} 
respectively).
Now let's perform $C \bigotimes A$.
The result is E, see Figure ~\ref{mat5}.

Recall that the result matrix is filled in by 6, 2 x 2 blocks, over the rows
and columns using the encoding discussed above. Notice how complicated the
indirect addressing becomes using this approach to implementation of the
Kronecker product. Notice also that if we wanted to distribute the computation
of a block of rows to 4 processors, we'd need multiple components of the left
argument.

\begin{figure*}[h]
\begin{center}
\includegraphics[width=7cm]{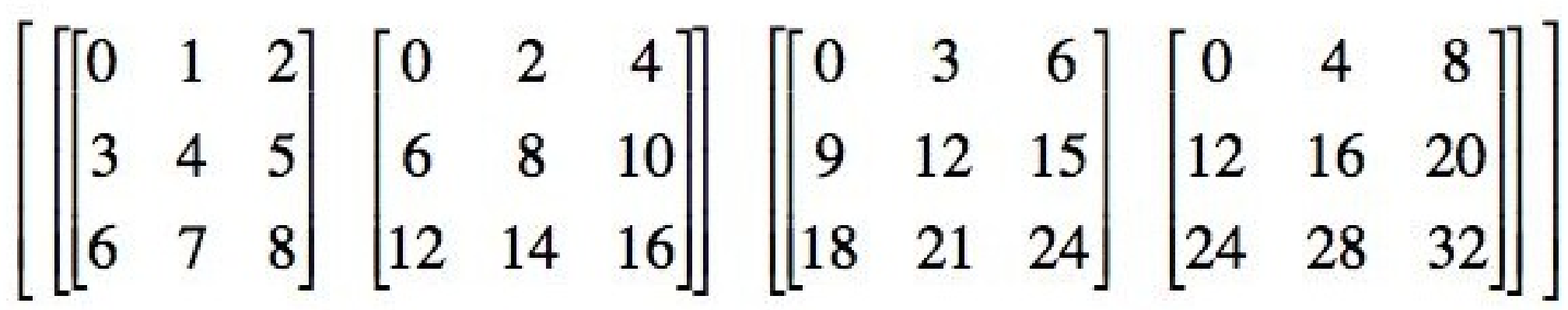}
\caption{\label{mat1}}
\end{center}
\end{figure*}

Let us now look at doing the same operations, i.e. multiple outer products, using
the MoA $\psi$ calculus approach, $C = A\, op_\times B$, as seen in Figure ~\ref{mat1},
%
%
%
%
%
is a 4-d array with shape 2 x 2 x 3 x 3. It is easy to see that
indexing this array with {\em partial indices} yields 3 x 3 sub-arrays. That is, the indices,
$<\!0\,0\!>$, $<\!0\,1\!>$, $<\!1\,0\!>$ and $<\!1\,1\!>$
are used to index C and each sub-array would be sent to available processors 
0-3, to create a start, stop, stride, mapping suitable for all architectures to date.

Now let's  perform $C\, op_\times A$. This would yield a 6-d array with shape
2 x 2 x 3 x 3 x 2 x 2. We can easily pull apart the arguments in the
operations. Let's now think of this array as a 4 x 3 x 3 x 2 x 2. We then
use the 4 to index the processors. We know the blocks have 36 components.

The following expressions illustrate how easy it is to compose, map,
and scale to a multi-processor architecture. We first get the shape.
\begin{eqnarray*}
\rho (( A \;op\; B ) \;op \;A) &=& (\rho ( A \;op\; B) \cat ( \rho A))\\
&=& (\rho A) \cat (\rho B) \cat (\rho A) \\
&=& < 2\; 2\; 3\; 3\; 2\; 2>
\end{eqnarray*}
The indices are composed as follows:
Given $ 0 \leq^* \; \; <\!i\;j\!> \;\;  <^* \;\; <\!2\;2\!>$;
\hfill
\break
$ 0 \leq^* \;\; <\!k\;l\!> \;\; <^* \;\; <\!3\;3\!>$; and
$ 0 \leq^* \;\; <\!m\;n\!> \;\; <^* \;\; <\!2\;2\!>$ 
 and 
\hfill \break
for all 
$ 0 \leq^* \;\; <\!i\;j\;k\;l\;m\;n\!> \;\; <^* \;\; <\!2\;2\;3\;3\;2\;2\!>$;
 \begin{eqnarray*}
 <i\;j\;k\;l\;m\;n> \psi \;((A \;op_\times \;B) \;op_\times \; A) &=& 
 ( <i\;j\;k\;l> \psi \;(A \;op_\times\;B)) \times (<m\;n> \psi A) \\
 &=& (<i\;j> \psi A) \;\times (<k\;l> \psi B) \;\times (<m\;n> \psi A) 
   \end{eqnarray*} 
   From here we can easily map chunks to the four processors using
   starts, stops, and strides.
   
   Let's take the above, referred to as the {\em Denotational
   Normal Form} (DNF) expressed in terms of {\em Cartesian
   coordinates} and transform it  into   its equivalent
   {\em Operational Normal Form}, (ONF), expressed in terms of start, stop, stride and count. The DNF is independent of layout. The ONF requires one. Let's assume row-major. 
   We'll see how natural that is for
   the Kronecker product at all levels of implementation.
    
  Let's break up the above multiple Kronecker product over 4 processors. We'll need to restructure the array's shape
   $<2\;2\;3\;3\;2\;2>$, to 
   $<4\;3\;3\;2\;2>$. This allows us to index the first dimension of this abstraction over the processors. We'll also index the first component of the leftmost argument by this value. Notice that the entire right argument is used/accessed in all of the processors. Thus, we think of the entire result of both products residing in an array with $\; \pi <4\;3\;3\;2\;2> \;= 144$ components (the $\pi$ operator
gives the product of the elements of the vector) laid out contiguously in memory using a row-major ordering.  
   
   Thus the equation above becomes for $0\leq p < 4$
     \begin{eqnarray*}
 <i\;j\;k\;l\;m\;n> \psi \;((A \;op_\times \;B) \;op_\times \; A) &=& 
 <p\;k\;l\;m\;n> \psi \;((\vec a \;op_\times \;B) \;op_\times \; A\\
 &=&( (<p> \psi \vec a) \;\times (<k\;l> \psi B)) \;\times (<m\;n> \psi A) 
   \end{eqnarray*} 
   
   The expression below describes what each processor, $p$, will
   do. $\vec a$ above denotes  the restructuring of $A$. avec and bvec are used to describe generic implementations.
   \noindent
   $\forall \; p,q,r\;\;s.t.\;0\leq p<4\; ; \;0\leq q <  9 ; \;\;
   0\leq r < 4$
   \begin{verbatim}
  ( avec[p] x bvec[q ]) x avec[r]
   \end{verbatim}
   We are able to collapse the 2-d indexing for A and B since their access is contiguous.
  This type of thinking and reasoning has been used for over 20 years\cite{mul88,Mul03,mullin-raynolds-book}.

  \section{Conclusion}
  The purpose of this paper was to illustrate how the Kronecker product/outer product is implemented, i.e. the algorithm used to represent the Kronecker/Tensor product, can hinder or exploit reasoning of resource management, performance, scalability, and portability of the algorithm. The classical way works but is  not easy to represent, compose, and partition over processor/memory hierarchies.
  
  MoA and Psi Calculus provide a way to reason about array based computing. By using shapes and the $\psi$ function to define a small algebra, higher order
  operations can be defined, composed, optimized, and mapped to a simple machine
  abstraction: {\em start, stop, stride, count}.
  
  Moving the theory to implementations that automatically generate correct optimal code is the next step. Over 20 years have been spent building prototypes to show proof of concept. Serious implementations must be initiated, studied,
  and advanced.
  
\newpage
\appendix

\section{Motivation for diadics, Kronecker and outer products}

This section provides some simple examples of how dyadics and Kronecker
products arise naturally in applied problems.

\subsection{Example from engineering}

In the field of electricity and magnetism the following operator arises
in the wave equation for the electric field:
\begin{equation}
\nabla \times \nabla \times {\vec E} = -\nabla^2 {\vec E} + \nabla (\nabla \cdot {\vec E}).
\label{operator}
\end{equation}
For a known source current density $\vec J({\vec r},t)$ (with a known
Fourier expansion) it is natural to expand the electric field in a 
Fourier expansion.  Thus we are let to consider the action of the 
operator of Eq.~\ref{operator} on a single Fourier component:
\begin{equation}
{\vec E}_{({\vec q},\omega)}({\vec r},t) = {\vec E}({\vec q},\omega) 
\exp(i({\vec q}\cdot{\vec r} - \omega t))
\label{component}
\end{equation}
Action on Eq.~\ref{component} with the operator of Eq.~\ref{operator}
gives:
\begin{equation}
q^2 {\vec E} - {\vec q}({\vec q} \cdot {\vec E}).
\end{equation}
This is simplified by introducing the {\em dyadic} (or Kronecker product):
${\vec q} \otimes {\vec q}$,
by writing:
\begin{equation}
q^2 {\vec E} - {\vec q}({\vec q} \cdot {\vec E}) \equiv
(q^2 {\hat I} - {\vec q} \otimes {\vec q}) \cdot {\vec E},
\label{diadeq}
\end{equation}
where $\hat I$ is the unit tensor ({\em matrix}) and the {\em dyadic}
${\vec q} \otimes {\vec q}$, is defined by its action on any other vector
$\vec u$ as follows:
\begin{equation}
({\vec q} \otimes {\vec q}) \cdot {\vec u}
\equiv
{\vec q} ({\vec q} \cdot {\vec u}).
\end{equation}
A convenient interpretation of the dyadic ${\vec q} \otimes {\vec q}$, arises
if we work with the unit vector ${\hat q} \equiv {\vec q}/q$, where $q$ is 
the magnitude of the vector $\vec q$.  In terms of the unit vector $\hat q$,
Eq.~\ref{diadeq}, becomes:
\begin{equation}
q^2 {\vec E} - {\vec q}({\vec q} \cdot {\vec E}) \equiv
q^2 ({\hat I} - {\hat q} \otimes {\hat q}) \cdot {\vec E},
\label{hateq}
\end{equation}
and we recognize the operator in parenthesis on the right hand side of 
Eq.~\ref{hateq} as a {\em projection operator}.

Indeed, the vector $({\hat q} \otimes {\hat q}) \cdot {\vec E}$ represents the 
component of $\vec E$ along the direction of $\hat q$ and the vector
$({\hat I} - {\hat q} \otimes {\hat q}) \cdot {\vec E}$ represents the 
component of $\vec E$ perpendicular to $\hat q$.  Explicitly we see
\begin{equation}
({\hat I} - {\hat q} \otimes {\hat q}) \cdot ({\hat q} \otimes {\hat q}) = 0.
\end{equation}
This follows from 
\begin{equation}
({\hat q} \otimes {\hat q}) \cdot ({\hat q} \otimes {\hat q}) 
= {\hat q} \otimes {\hat q},
\end{equation}
which is a natural consequence of the above definitions.

Another contribution to the equation for electromagnetic waves in 
an anisotropic medium is the displacement field $\vec D$ that is related
to the electric field $\vec E$ by the equation:
\begin{equation}
{\vec D} = {\bar \epsilon} \cdot {\vec E},
\end{equation}
where $\bar \epsilon$ is the {\em dielectric tensor} with diagonal components
$\bar \epsilon_{||}$ and off-diagonal components $\bar \epsilon_\perp$.  The
components $\bar \epsilon_{||}$ and $\bar \epsilon_\perp$ arise from the fact
that the response of the medium is different for electric field components
parallel to, and perpendicular to the direction of wave propagation $\hat q$,
respectively.

Using the dyadic notation, the dielectric tensor $\bar \epsilon$ is 
conveniently written as
\begin{equation}
{\bar \epsilon} = {\bar \epsilon_{||} } ({\hat q} \otimes {\hat q}) 
+ {\bar \epsilon_\perp} ({\hat I} - {\hat q} \otimes {\hat q}). 
\end{equation}
The complete wave equation, in Fourier space, reads:
\begin{equation}
{\bar A}({\vec q},\omega) \cdot {\vec E} ({\vec q},\omega) = {\vec b} ({\vec q},\omega)
\label{waveeq}
\end{equation}
where the operator
${\bar A}({\vec q},\omega)$ is the sum of the {\em longitudinal} component:
\begin{equation}
{\bar A}_{||}({\vec q},\omega) = {\frac {\omega^2 {\bar \epsilon_{||} ({\vec q},\omega) } }{c^2}}
({\hat q} \otimes {\hat q}),
\end{equation}
and the {\em transverse} component:
\begin{equation}
{\bar A}_\perp({\vec q},\omega) = \Biggl ({\frac {\omega^2 {\bar \epsilon_\perp ({\vec q},\omega) } }{c^2} - q^2} \Biggr )
({\hat I} - {\hat q} \otimes {\hat q}).
\end{equation}
The right hand side of the wave equation (Eq.~\ref{waveeq}) is defined in
terms of the known {\em current density} ${\vec J} ({\vec q},\omega)$ 
by:
\begin{equation}
{\vec b} ({\vec q},\omega) \equiv {\frac {4\pi i \omega}{c^2}} {\vec J} ({\vec q},\omega)
\end{equation}

The wave equation (Eq.~\ref{waveeq}) is solved through the use of the 
{\em dyadic Green's function} ${\bar G} ({\vec q},\omega)$
\begin{equation}
{\vec E} ({\vec q},\omega) = 
{\vec G} ({\vec q},\omega) \cdot
{\vec b} ({\vec q},\omega),
\label{invert}
\end{equation}
where ${\vec G} ({\vec q},\omega)$ is the inverse of 
${\vec A} ({\vec q},\omega)$ and thus satisfies:
\begin{equation}
{\vec G} ({\vec q},\omega) \cdot {\vec A} ({\vec q},\omega) = {\hat I}.
\end{equation}
The {\em longitudinal} and {\em transverse} components of the dyadic Green's
function are given explicitly by:

\begin{equation}
{\bar G}_{||}({\vec q},\omega) = 
{ \frac
{({\hat q} \otimes {\hat q})}
{\Bigl ({\frac {\omega^2 {\bar \epsilon_{||} ({\vec q},\omega) } }{c^2}}\Bigr )}
},
\end{equation}
and,
\begin{equation}
{\bar G}_\perp({\vec q},\omega) = {\frac {({\hat I} - {\hat q} \otimes {\hat q})} {\Biggl ({\frac {\omega^2 {\bar \epsilon_\perp ({\vec q},\omega) } }{c^2} - q^2} \Biggr )}}.
\end{equation}
Thus a complete solution of the original problem is obtained by Fourier
transforming Eq.~\ref{invert}.

\subsection{Example from linear algebra: matrix decompositions}

Kronecker products ({\em dyadics}) can also be conveniently used to express
a matrix expansion.  Consider a Hermitian matrix $H$ and its normalized 
eigenvectors ${\vec u}_j$ (i.e. ${\vec u_i} {\vec u_j} = \delta_{ij}$) and 
eigenvalues $\lambda_j$ satisfying $H u_j = \lambda_j u_j$.

A well-known result of linear algebra is that the matrix $H$ can be 
expressed in terms of the following expansion involving Kronecker products:
\begin{equation}
H = \lambda_1 {\vec u_1} \otimes {\vec u_1}, + \lambda_2 {\vec u_2} \otimes {\vec u_2} + \cdots.
\label{expansion}
\end{equation}
This expansion follows from the fact that the eigenvectors form a complete
basis and, as such, any arbitrary vector can be expanded as a sum of the
eigenvectors as:
\begin{equation}
{\vec v} = c_1 {\vec u_1} + c_2 {\vec u_2} + \cdots
\label{vexpand}
\end{equation}
We see again, the natural interpretation of the Kronecker products as 
{\em projection operators}.  Each term in the expansion of Eq.~\ref{expansion}
gives a non-zero result only when acting on the corresponding eigenvector of
Eq.~\ref{vexpand}.
The result, $\lambda_j c_j {\vec u_j}$, is identical to the action of 
$H$ acting on the corresponding component $c_1 {\vec u_1}$ in the vector
expansion of Eq.~\ref{vexpand}.
In other words, we find  
\begin{equation}
({\vec u_i} \otimes {\vec u_i}) \cdot {\vec u_j} = \delta_{ij} {\vec u_i}.
\end{equation}
Thus, this section is consistent with the previous section in terms of the
interpretation of Kronecker products as projection operators.

\subsection{Higher dimensional generalizations}

We now consider an example from the theory of orthogonal functions (i.e. 
Hilbert space).  For this discussion, it is convenient to use Dirac notation.
We expand a given function $|\psi\!>$ in a complete set of basis functions
$|\ell,m\!>$ as:
\begin{equation}
|\psi\!> = \sum_{\ell, m} C_{\ell,m} |\ell,m\!>.
\label{cexpand}
\end{equation}
By orthogonality we see that the coefficients can be written in terms 
of an inner product:
\begin{equation}
C_{\ell,m} = <\!\ell,m |\psi\!>,
\end{equation}
which is interpreted as a {\em projection} onto the basis vector (function)
$|\ell,m\!>$. 
We now wish to expand in another complete basis set $|\ell^\prime,m^\prime\!>$, perhaps obtained from the starting set by rotating the coordinate system
\begin{equation}
|\psi\!> = \sum_{\ell^\prime, m^\prime} B_{\ell^\prime,m^\prime} |\ell^\prime,m^\prime\!>.
\label{bexpand}
\end{equation}
The coefficients of this expansion are likewise expressed as:
\begin{equation}
B_{\ell^\prime,m^\prime} = <\!\ell^\prime,m^\prime |\psi\!>,
\end{equation}
and we can relate the coefficients of this later expansion to the coefficients
the former expansion by taking the inner product of Eq.~\ref{bexpand}
with the basis function $|\ell,m\!>$, to yield:
\begin{equation}
<\!\ell,m |\psi\!> = \sum_{\ell^\prime,m^\prime} <\ell,m|\ell^\prime,m^\prime\!> <\ell^\prime,m^\prime|\psi>,
\label{exp1}
\end{equation}
which can more simply be written:
\begin{equation}
C_{\ell,m} = \sum_{\ell^\prime,m^\prime} <\ell,m|\ell^\prime,m^\prime\!>  
B_{\ell^\prime,m^\prime}
\label{exp2}
\end{equation}
From equations~\ref{exp1} and~\ref{exp2} we see the natural definition of 
the unit operator:
\begin{equation}
{\hat I} = 
\sum_{\ell^\prime,m^\prime} |\ell^\prime,m^\prime\!> <\ell^\prime,m^\prime|
\end{equation}
Note the close analogy between this expansion and the expansion of the 
Hermitian matrix in terms of its eigenvectors in Eq.~\ref{expansion}.  We
see therefore, that a higher-dimensional analog of Eq.~\ref{expansion}
would be the operator expansion of
\begin{equation}
{\hat L} = 
\sum_{\ell^\prime,m^\prime} \lambda_{\ell,m} |\ell^\prime,m^\prime\!> <\ell^\prime,m^\prime|,
\label{higher}
\end{equation}
where $\lambda_{\ell,m}$ and $|\ell,m\!>$ are the eigenvalues and eigenfunctions
of the operator $\hat L$, respectively (i.e. ${\hat L} |\ell,m\!>  = 
\lambda_{\ell,m}|\ell,m\!>$).

If we express operators such as Eq.~\ref{higher} in matrix form, we are
naturally led to a higher dimensional generalization of the dyadic
${\vec u} \otimes {\vec u}$, namely, the matrix product: 
${\bar A} \otimes {\bar A}$, where $\bar A$ is a matrix or higher dimensional 
tensor.  Compositions of such products, such as ${\bar A} \otimes {\bar B} 
\otimes {\bar C}$ are also similarly defined.
\subsection{Matrix form}
The matrix representation of the Kronecker Product is
\begin{equation}
({\bar A} \otimes {\bar B})_{I,J} = A_{i,j} B_{\ell,m} 
\end{equation}
where $I \equiv <i,\ell>$ and $J = <j,m>$ are composite indices that 
cycle through the integers as $<i,\ell>$ and $<j,m>$ cycle through their
allowed values in row major order (i.e. $\ell$ cycles faster than $i$ and
$m$ cycles faster than $j$), and the eigenvector $U_J$ is constructed 
as a large column vector with as many copies of the eigenvector $\vec u_i$
of $B$ as there are columns of $A$.

\bibliography{paper}

\end{document}